\documentclass[twocolumn,prl]{revtex4-1}
\usepackage{amsfonts}
\usepackage[T1]{fontenc}
\usepackage{amsmath,amsbsy,amssymb,graphicx}
\usepackage{times}
\usepackage{amsmath}
\usepackage{amssymb}
\usepackage{graphicx}

\begin{document}

\title{{\Large Quantum Percolation Transition from Graphene to Graphane:}\\
{\Large Graph Theoretical Approach}}
\author{Motohiko Ezawa}
\affiliation{Department of Applied Physics, University of Tokyo, Hongo 7-3-1, 113-8656,
Japan }

\begin{abstract}
Graphane is obtained by perfectly hydrogenating graphene. There exists an
intermediate material, partially hydrogenated graphene (which we call 
\textit{hydrographene}), interpolating from pure graphene to pure graphane.
It has various intriguing electronic and magnetic properties. We investigate
a metal-insulator transition, employing a quantum-site percolation model
together with a graph theoretical approach. Hydrographene is an exceptional
case in which electronic properties cannot be determined solely by the
density of states at the Fermi energy. Though there are plenty of zero
energy state in wide range of hydrogenation density, most of them are
insulating states. We also demonstrate that it shows a bulk ferromagnetic
property based on the Lieb theory.
\end{abstract}

\date{\today }
\maketitle

Graphene is one of the main topics in condensed matter physics because of
its unusual electronic properties\cite{GraphEx,NetoRev}. It is a promising
material to design future nanoelectronic devices. To make a device it is
necessary to confine electrons within a certain finite domain. However, this
is impossible in a graphene sheet by applying electric field due to the
Klein tunneling effect, which is an intrinsic nature of Dirac electrons. One
method is to cut graphene to form a required domain. Resultant graphene
derivatives such as graphene nanoribbons\cite{Ribbon} and graphene nanodisks%
\cite{Disk} show remarkable electronic and magnetic properties.

\begin{figure}[t]
\centerline{\includegraphics[width=0.5\textwidth]{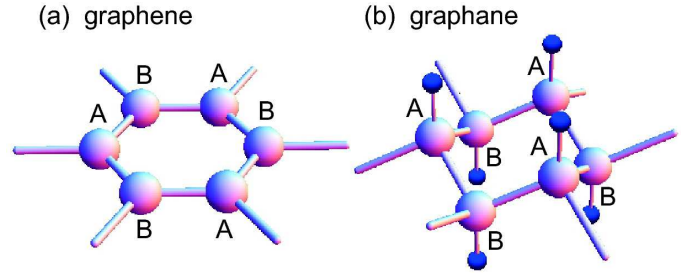}}
\caption{{}(Color online) (a) Illustration of graphene. There are carbons on
a honeycomb lattice. They are grouped in two inequivalent sites A and B. (b)
Illustration of graphane. Hydrogens are attached upwardly to A sites, and
downwardly to B sites. A honeycomb lattice is distorted.}
\label{FigGraphane}
\end{figure}

Recently graphane has been attracting much attention\cite%
{Sofo,Elias,Balog,Savche}. It is a graphene derivative obtained by perfectly
hydrogenating graphene (Fig.\ref{FigGraphane}). Graphene is a semimetal with
each carbon forming an sp$^{2}$ orbital, while graphane is an insulator with
each carbon forming an sp$^{3}$ orbital. Graphane provides us with an
alternative method of confining electrons within a finite domain, since $\pi 
$-electrons are excluded from hydrogenated carbons. Namely, by
dehydrogenating a finite region of a graphane sheet, electron can be
confined within this region. By dehydrogenating a graphane sheet according
to a circuit pattern, in principle it is possible to fabricate a
graphene-graphane complex, embodying various electronic functions\cite%
{NovPhys}.

In this paper, we explore an intermediate material, \textit{partially
hydrogenated graphene}, which interpolates from pure graphene to pure
graphane\cite{Balog}. Let us call it \textit{hydrographene}. There are two
types of hydrogenation, single-sided percolation and double-sided
percolation. Employing a graph theoretical approach to the site-percolation
model\cite{PercoBook}, we present an intuitive and physical picture
revealing the electronic and magnetic properties of hydrographene with
hydrogenation density $q$, $0\leq q\leq 1$. As $q$ increases, there occurs a
metal-insulator transition at $q_{\text{c}}$. It is mapped to a
ferromagnet-paramagnet transition in a Potts model with $q_{\text{c}}$
corresponding to a critical temperature $T_{\text{c}}$. We remark that
hydrographene is an exceptional case in which electronic properties cannot
be determined solely by the density of states (DOS) at the Fermi energy.
This is because there emerge a number of insulating states which contribute
to the DOS at the Fermi energy but not to the conductivity. We also show
that it is a bulk ferromagnet in the context of the Lieb theory. We find
that single-sided hydrogenation is more efficient to form large magnetic
moment than double-sided hydrogenation.

\textit{Model:} In hydrographene a finite density of hydrogens are absorbed
randomly to graphene\cite{Balog}. Carbons with absorbed hydrogen form an sp$%
^{3}$ bond, where no $\pi $-electron exists. To simulate this fact we model
the system by the Hamiltonian,%
\begin{equation}
H=\sum_{\left\langle i,j\right\rangle }tc_{i}^{\dagger }c_{j}+V{\sum_{r}}%
^{\prime }c_{r}^{\dagger }c_{r},
\end{equation}%
where $c_{j}$ ($c_{i}^{\dagger }$) is the annihilation (creation) operator
of $\pi $-electron, $t=2.7$ [eV] is the transfer energy, and $\left\langle
i,j\right\rangle $ denotes the next-nearest neighbor sites of the honeycomb
lattice. To exclude the electron at the hydrogenated site $r$ we have
imposed an infinitely large on-site potential $V$ at $r$. Thus the sum ${%
\sum_{r}^{\prime }}$ runs over the lattice sites where carbons are
hydrogenated.

We analyze a quantum site-percolation problem based on this Hamiltonian.
Electromagnetic properties depend on percolation networks (Fig.\ref%
{FigHoneyGraph}), where hydrogenated carbons have been removed from the
network. The resultant carbon network is nothing but a graph with vertices
corresponding to carbons. The model Hamiltonian has one-to-one
correspondence to the adjacency matrix of a graph. Though the Hamiltonian is
noninteracting, the problem is highly nontrivial because percolation network
is highly nontrivial.

Graphene is a bipartite system made of carbons at two inequivalent sites (A
and B sites) on a honeycomb lattice, as illustrated in Fig.\ref{FigGraphane}%
(a). These carbons can absorb hydrogens randomly. The reaction with hydrogen
is reversible, so that the original metallic state and the lattice spacing
can be restored by annealing\cite{Elias}. However, the way of attachment is
opposite between the A and B sites in graphane [Fig.\ref{FigGraphane}(b)].
As a matter of convenience, hydrogens are absorbed upwardly to A sites and
downwardly to B sites. We can investigate two types of percolation problems
in hydrographene. In one case, hydrogenation occurs randomly on both A and B
sites. We call it double-sided percolation. In the other case, hydrogenation
occurs only on A sites. We call it single-sided percolation. Single-sided
hydrogenation can be manufactured by resting graphene on a Silica surface%
\cite{Elias}.

The number of lattice sites with no defects is $N_{\text{c}}$. We assume $N_{%
\text{c}}$ is even for simplicity. We define the hydrogenation density by $%
q=M/N_{\text{c}}$ for double-sided hydrogenation and $q=2M/N_{\text{c}}$ for
single-sided hydrogenation, where $M$ is the number of hydrogenated carbons, 
$M={\sum_{r}^{\prime }}$. We choose the positions of hydrogenated carbons by
the Monte Carlo method. Physical quantities are to be determined by taking
the statistical average.

\begin{figure}[t]
\centerline{\includegraphics[width=0.5\textwidth]{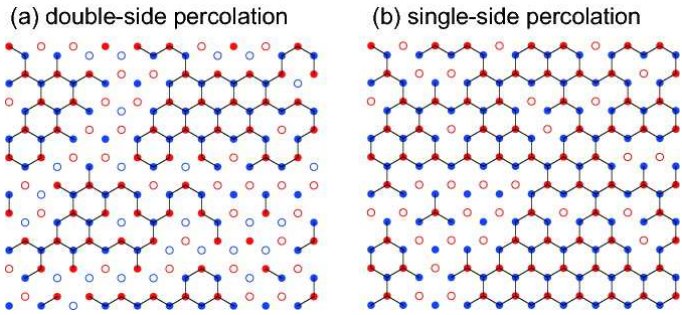}}
\caption{{}(Color online) Graph representation of percolation network in a
honeycomb lattice. (a) 30\% hydrogenation on the two sides. The system is
not connected. (b) 30\% hydrogenation on one side. The system is connected. }
\label{FigHoneyGraph}
\end{figure}

\textit{Isolated carbons:} We refer to carbons with no adjacent carbons as
isolated carbons, to those with one adjacent carbon as edge carbons, to
those with two adjacent carbons as corner carbons, and to those with three
adjacent carbons as bulk carbons. We apply a combinatorial theory and
estimate the numbers of these different types of carbons by using the fact
that each site is occupied with probability $q$. We also use the quantity $%
p=1-q$ in what follows.

\begin{figure}[t]
\centerline{\includegraphics[width=0.5\textwidth]{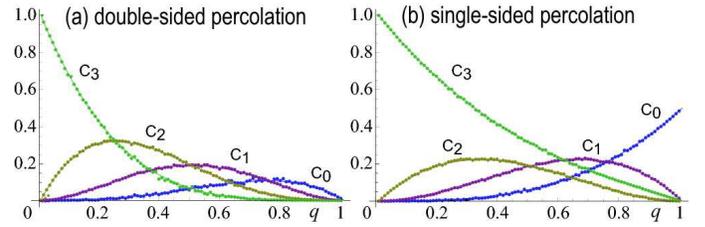}}
\caption{{}(Color online) The numbers of isolated carbons $C_{3}$, edge
carbons $C_{2}$, corner carbons $C_{1}$ and bulk carbons $C_{0}$ in unit of $%
N_{\text{c}}$. The horizontal axis is the hydrogenation density $q$.}
\label{FigEdge}
\end{figure}

In the case of double-sided percolation, we calculate the number of isolated
carbons to be $C_{0}=pq^{3}N_{\text{c}}$, that of edge carbons to be $%
C_{1}=3p^{2}q^{2}N_{\text{c}}$, that of corner carbons to be $%
C_{2}=3p^{3}qN_{\text{c}}$ and that of bulk carbons to be $C_{3}=p^{4}N_{%
\text{c}}$. They satisfy the relation $C_{0}+C_{1}+C_{2}+C_{3}=pN_{\text{c}}$%
. The total bond number is $\frac{1}{2}\left( C_{1}+2C_{2}+3C_{3}\right) =%
\frac{3}{2}p^{2}N_{\text{c}}$.

In the case of single-sided percolation, we calculate the number of isolated
carbons to be $C_{0}=\frac{1}{2}q^{3}N_{\text{c}}$, that of edge carbons to
be $C_{1}=\frac{3}{2}pq^{2}N_{\text{c}}$, that of corner carbons to be $%
C_{2}=\frac{3}{2}p^{2}qN_{\text{c}}$ and that of bulk carbons to be $C_{3}=%
\frac{1}{2}\left( p^{3}+p\right) N_{\text{c}}$. They satisfy the relation $%
C_{0}+C_{1}+C_{2}+C_{3}=\frac{1}{2}\left( p+1\right) N_{\text{c}}$. The
total bond number is $\frac{1}{2}\left( C_{1}+2C_{2}+3C_{3}\right) =3pN_{%
\text{c}}$.

We show the numbers of various types of carbons estimated in this way in Fig.%
\ref{FigEdge}. In order to verify that the combinatorial theory yields
correct results, we have also carried out numerical calculations, and found
that the results agree completely between them as in Fig.\ref{FigEdge}.

\textit{Zero-energy states:} We investigate the number of the zero-energy
state $N_{0}$. It is given by diagonalizing the Hamiltonian, or equivalently
by the difference between the dimension and the rank of the Hamiltonian, $%
N_{0}=\dim H-$rank$H$. The analysis is quite different between double-sided
and single-sided percolations.

\begin{figure}[t]
\centerline{\includegraphics[width=0.5\textwidth]{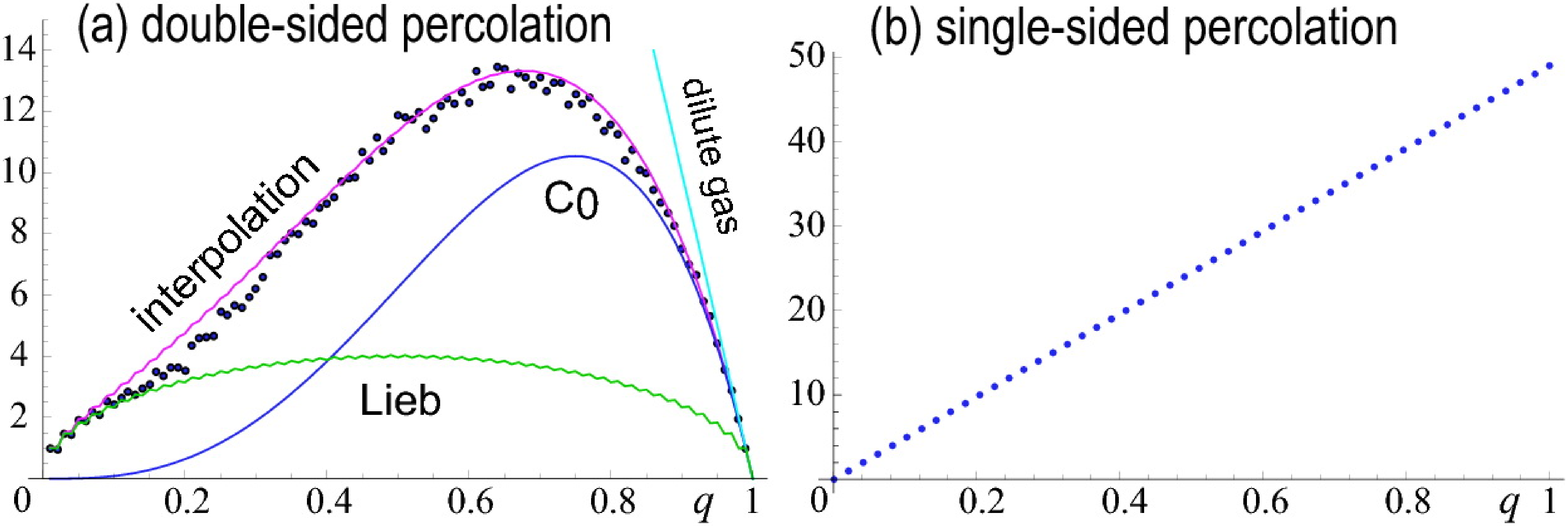}}
\caption{{}(Color online) The number of the zero energy states. The unit
cell size is 100. The average is taken 1000 times. }
\label{FigNZero}
\end{figure}

We first discuss the double-sided case. We show the numerical results in Fig.%
\ref{FigNZero}(a), where we have adopted the periodic boundary condition to
the unit cell with sizes $100$. We have also given the number of isolated
carbons in the same figure. Note that one isolated carbon yields one zero
mode. It is observed that most part of the zero modes have arisen from
isolated carbons for high density hydrogenation limit ($q\approx 1$). A
comment is in order. A metallic conductivity is proportional to $N_{0}$.
However, this is not the case in hydrographene since isolated carbons cannot
carry electric charges. We expect a metal-insulater transition to occur,
about which we give a detailed discussion soon after in this paper.

We give physical interpretations of the zero modes. In the low density
hydrogenation limit ($q\approx 0$), all sites are connected, i. e. the
cluster number is only one. We can apply the Lieb theorem, which states that
the number of the zero energy states is determined by the difference of the
A-site and B-site. It is exactly obtained as%
\begin{equation}
N_{0}^{\text{L}}=\sum_{s=0}^{M}\left\vert M-2s\right\vert \left( 
\begin{array}{c}
N_{\text{c}}/2 \\ 
s%
\end{array}%
\right) \left( 
\begin{array}{c}
N_{\text{c}}/2 \\ 
M-s%
\end{array}%
\right) \left( 
\begin{array}{c}
N_{\text{c}} \\ 
M%
\end{array}%
\right) ^{-1}
\end{equation}%
in terms of binomial coefficients. The distribution is symmetric at $M=N_{%
\text{c}}/2$. We have shown $N_{0}^{\text{L}}$ by the choice of $N_{\text{c}%
}=100$ in Fig.\ref{FigNZero}(a). By taking $N_{\text{c}}\rightarrow \infty $%
, we obtain%
\begin{equation}
N_{0}^{\text{L}}=\sqrt{2N_{\text{c}}pq/\pi }.  \label{EqLieb}
\end{equation}%
On the other hand, in the high density hydrogenation limit ($q\approx 1$),
we can apply the dilute gas model. Every sites are isolated or unconnected
with each other and give zero energy states. Then the number of zero is
simply given by%
\begin{equation}
N_{0}^{\text{D}}=N_{\text{c}}(1-q).
\end{equation}%
We have shown $N_{0}^{\text{D}}$ by the choice of $N_{\text{c}}=100$ in Fig.%
\ref{FigNZero}(a).

A simple interpolating formula reads as

\begin{equation}
N_{0}\simeq C_{0}+\frac{1}{6}C_{1}+\frac{C_{2}+C_{3}}{C_{0}+C_{1}+C_{2}+C_{3}%
}N_{0}^{L},
\end{equation}%
where $C_{0}$, $C_{1}$, $C_{2}$ and $C_{3}$ are the numbers of isolated
carbons, edge carbons, corner carbons and bulk carbons, respectively. It
explains the numerical result reasonably well, as in Fig.\ref{FigNZero}(a).
The first and second terms represent the contributions from the isolated and
edge carbons, respectively. The third term follows from the Lieb theorem
with an appropriate correction.

We now discuss the zero-energy states in single-sided percolation. The
number of the zero-energy states in each cluster is given by $N_{0}^{\text{%
cluster}}=\left\vert N_{B}^{\text{cluster}}-N_{A}^{\text{cluster}%
}\right\vert $ in general. Here, $N_{B}^{\text{cluster}}>N_{A}^{\text{cluster%
}}$ since only A sites are hydrogenated. Hence, the total number of the
zero-energy states is given by 
\begin{equation}
N_{0}=\sum N_{0}^{\text{cluster}}=N_{B}-N_{A}=qN_{\text{c}}.  \label{EqS}
\end{equation}%
This is confirmed perfectly by the numerical result in Fig.\ref{FigNZero}%
(b). Thus, the numbers of the zero-energy states are very different between
one-sided and double-sided hydrogenations. 
\begin{figure}[t]
\centerline{\includegraphics[width=0.5\textwidth]{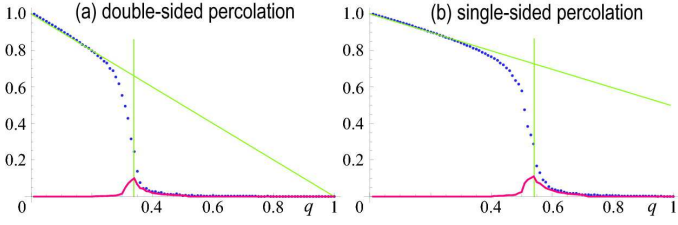}}
\caption{{}(Color online) The maximum cluster size (black dots) and the
second maximum cluster size (red curves) for various hydrogenation density $%
q $ in unit of $N_{c}$. The unit cell size is 10000. The average is taken
100 times.}
\label{FigMaxSize}
\end{figure}

\textit{Metal-insulator transition:} The bulk electronic properties of
hydrographene is closely related to the connectivity of graph. We have
calculated numerically the maximum cluster size $N^{\text{mc}}$ as a
function of hydrogenation in the system with $N_{\text{c}}=10^{4}$, which we
show in Fig.\ref{FigMaxSize}. The ratio $N^{\text{mc}}/N_{\text{c}}$ is
approximately equals to the percolation probability $P(q)$. It is by
definition the probability for an infinitely large cluster to appear in an
infinitely large system. In the present system it may be interpreted as a
probability that the right-hand side and the left-hand side of a
hydrographene sheet is connected by a cluster.

It is seen in Fig.\ref{FigMaxSize} that the maximum cluster size decreases
almost linearly as $q$ increases, and suddenly becomes zero at a critical
density $q_{c}$. This is a characteristic feature of a phase transition.
Indeed, it is a phase transition since the present percolation system is
mapped to the ferromagnet as we point out later: See (\ref{KPmap}). For
double-sided percolation the critical density is determined as $q_{c}=0.32$,
which is consistent with the well-known result\cite{PercoBook} on
percolation in the honeycomb lattice. For single-sided percolation the
critical density is determined as $q_{c}=0.55$.

The percolation probability is closely related to the conductance of
hydrographene. For $q<q_{c}$, there is a large cluster around the sample.
Thus the hydrographene is metallic. On the other hand, the hydrographene is
insulator for $q>q_{c}$ because there is no such cluster. Namely, a
percolation-induced metal-insulator transition occurs at $q=q_{c}$. It is
intriguing that there is a large amount of the zero-energy states even
though $q>q_{c}$. This is highly contrasted with the normal metal, where the
metal or insulator can be determined by the existence of the zero-energy
states. In hydrographene, the connectivity of carbon network plays a crucial
role for the metal-insulator transition. 
\begin{figure}[t]
\centerline{\includegraphics[width=0.5\textwidth]{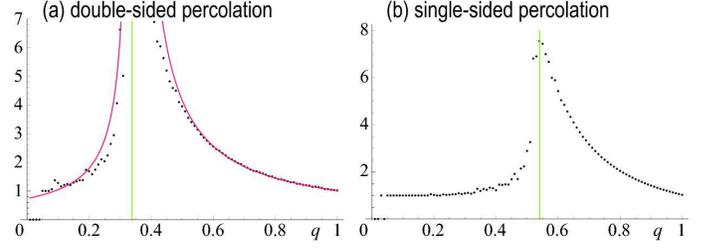}}
\caption{(Color online) The mean finite cluster size for various hydrogenation density $q$%
. The unit cell size is 40000. The average is taken 100 times.}
\label{FigSus}
\end{figure}

We have also calculated numerically the mean size of finite cluster $S\left(
q\right) $ as a function of hydrogenation, which we show in Fig.\ref{FigSus}%
. It diverge at $q_{c}=0.31$, indicating a critical behavior%
\begin{equation}
S\left( p\right) \varpropto \left\vert q-q_{c}\right\vert ^{-1}
\label{Curie}
\end{equation}%
for double-sided hydrogenation.

\begin{figure}[t]
\centerline{\includegraphics[width=0.5\textwidth]{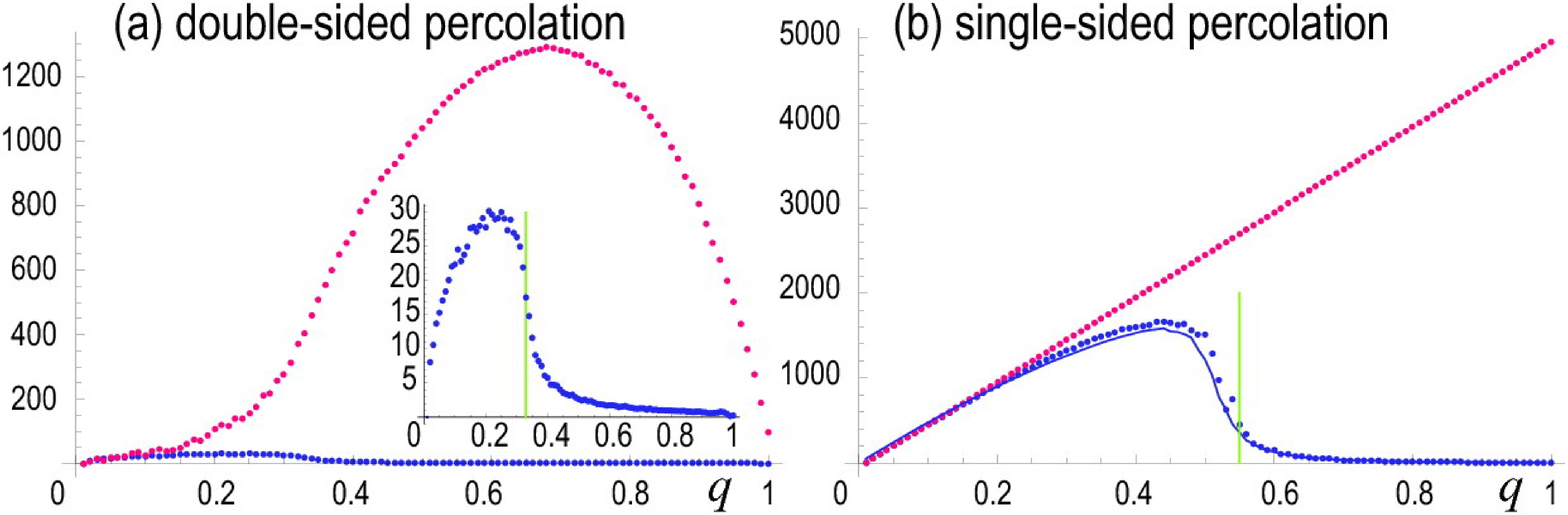}}
\caption{{}(Color online) The magnetization in unit of $\hbar /2$ in the
presence (red dots) and absence (blue dots) of ferromagnetic coupling for
various hydrogenation density $q$. A black curve represents the formula (%
\protect\ref{EqM}). Inset: An extended figure of the magnetization without
ferromagnetic coupling. The unit cell size is 10000. The average is taken
500 times.}
\label{FigAB}
\end{figure}

\textit{Magnetization:} We proceed to analyze the magnetic property of
hydrographene. We are able to make a general argument with respect to the
magnetic moment. According to the Lieb theorem valid to the bipartite system
at half-filling in connected graph, the magnetization $M$ is determined by
the difference between the numbers $N_{A}$ and $N_{B}$ of the A and B sites
in each cluster, $M=\frac{1}{2}\left\vert N_{A}-N_{B}\right\vert $, where
the spin direction is arbitrary in general. The magnetization of
hydrographene is determined by that of the largest cluster, since it is
dominant: See Fig.\ref{FigMaxSize}. We show the numerically calculated
magnetization as a function of $q$ in Fig.\ref{FigAB}. It is well
approximated by the relation 
\begin{equation}
M(q)=P\left( q\right) N_{0}^{\text{L}},  \label{EqM}
\end{equation}%
where $N_{0}^{\text{L}}$ is given by eq.(\ref{EqLieb}) for double-sided
hydrogenation and eq.(\ref{EqS}) for single-sided hydrogenation: See Fig.\ref%
{FigAB}(b). The system is ferromagnet for $q<q_{c}$, and paramagnet for $%
q>q_{c}$.

The maximum magnetization per site is obtained $M=3\times 10^{-3}$ at $%
q=0.21 $ for double-sided hydrogenation and $M=0.17$ at $q=0.44$ for
single-sided hydrogenation. Single-sided hydrogenation is about 500 times
more efficient than double-sided hydrogenation because only A sites are
hydrogenated in single-sided hydrogenation. The magnetization is
proportional to the number of the sites. In other words, hydrographene shows
the bulk magnetism, which is highly contrasted with the edge magnetism in
graphene nanoribbons and nanodisks, where magnetization is proportional to
the number of carbons along the edge. Bulk magnetism is desirable because
edge magnetism disappears in the thermodynamic limit.

A comment is in order. It is intriguing that the magnetization is zero for
the perfect single-sided hydrographene ($q=1$), which is in contradiction
with the previous result\cite{Zhou} obtained based on a density functional
theory study. This is not surprising since the direction of magnetization is
arbitrary in each cluster in our noninteracting model, as we have noted. The
spin directions of two adjacent clusters may be aligned due to an exchange
interaction if it is present, and it is probable that all clusters have the
same spin direction. The maximum magnetization per site is obtained $M=0.13$ at $%
q=0.68 $ for double-sided hydrogenation: See Fig.\ref{FigAB}(a). On the other hand, the magnetization increases
linearly as a function of the hydrogenation parameter $q$ for single-sided
hydrogenation: See Fig.\ref{FigAB}(b). This reproduces the result\cite{Zhou}
for the perfect single-sided hydrographene ($q=1$).

\textit{Percolation and ferromagnet:} There exists a close relation between
the percolation problem and the ferromagnet. Indeed, the percolation problem
is mapped to the zero-state Potts model on Kagom\'{e} lattice via the
Kasteleyn-Fortuin mapping\cite{KF}. The hydrogenation parameter $q$
corresponds to the temperature $T$ of a ferromagnet, 
\begin{equation}
q=\exp \left( -J/kT\right) ,  \label{KPmap}
\end{equation}%
with $J$ being the exchange stiffness. The percolation probability $P\left(
q\right) $ is mapped to the magnetization as a function of temperature $T$.
Thus, the metal-insulator transition corresponds to the
ferromagnet-paramagnet transition. Furthermore, the critical behavior (\ref%
{Curie}) corresponds to the Curie-Weiss law of the susceptibility. We
anticipate various properties familiar in ferromagnet to be translated into
hydrographene via the Kasteleyn-Fortuin mapping.

In summary, we have discovered various intriguing electronic and magnetic
properties of hydrographene based on a quantum-percolation model. We have
proposed a new-type of percolation, single-sided percolation, which is
qualitatively different from the usual percolation on honeycomb lattice.
Hydrographene is an ideal system to investigate from the viewpoint of
quantum site-percolation transition.

I am very much grateful to N. Nagaosa for fruitful discussions on the
subject. This work was supported in part by Grants-in-Aid for Scientific
Research from the Ministry of Education, Science, Sports and Culture No.
22740196.

\end{document}